\documentclass[showpacs,preprintnumbers]{revtex4}
\usepackage{graphicx}
\usepackage{color}

\def\bc{\begin{center}}
\def\ec{\end{center}}

\def\beq{\begin{equation}}
\def\eeq{\end{equation}}

\begin{document}

\title{On Bose-Einstein
condensation and superfluidity of trapped photons with coordinate-dependent mass and interactions}

\author{Oleg L. Berman$^{1,2}$, Roman Ya. Kezerashvili$^{1,2}$,  and Yurii E.
Lozovik$^{3,4}$}
 \affiliation{\mbox{$^{1}$Physics Department, New York
City College
of Technology, The City University of New York,} \\
Brooklyn, NY 11201, USA \\
\mbox{$^{2}$The Graduate School and University Center, The
City University of New York,} \\
New York, NY 10016, USA \\
\mbox{$^{3}$Institute of Spectroscopy, Russian Academy of Sciences,
142190 Troitsk, Moscow, Russia }\\
\mbox{$^{4}$National Research University Higher School of Economics,
Moscow, Russia}}

\date{\today}

\begin{abstract}
The condensate density profile of trapped two-dimensional gas of
photons in an optical microcavity, filled by a  dye solution,  is
analyzed taking into account a coordinate-dependent effective mass
of cavity photons and photon-photon coupling parameter. The profiles
for the densities of the superfluid and normal phases of trapped
photons in the different regions of the system at the fixed
temperature are analyzed. The radial dependencies of local
mean-field phase transition temperature $T_c^0 (r)$ and local
Kosterlitz-Thouless transition temperature $T_c (r)$ for trapped
microcavity photons are obtained.   The coordinate dependence of
cavity photon effective mass and photon-photon coupling parameter is
important  for the mirrors of smaller radius with the high trapping
frequency, which provides BEC and superfluidity for smaller critical
number of photons at the same temperature.
 We discuss a possibility of an experimental study of the density profiles
for the normal and superfluid components  in the system under consideration.

\end{abstract}


\maketitle

\section{Introduction}

\label{intro}

When a system of bosons is cooled to low temperatures, a substantial
fraction of the particles spontaneously occupy the single lowest
energy quantum state. This phenomenon is known as Bose-Einstein
condensation (BEC) and its occurs in many-particle systems of bosons
with masses $m$ and temperature $T$ when the de Broglie wavelength
of the Bose particle exceeds the mean interparticle
distance~\cite{Pitaevskii}. The most remarkable consequence of BEC
is that there should be a temperature below which a finite fraction
of all the bosons ``condense'' into the same one-particle state with
macroscopic properties described by a single condensate
wavefunction, promoting quantum physics to classical time- and
length scales.

Most recently, the observations at room temperature of the BEC of
two-dimensional photon gas confined in an optical microcavity,
formed by spherical mirrors and filled by a  dye solution,  were
reported~\cite{Klaers_Nature,Klaers_Nf,Nyman,natcom}. The
interaction between microcavity photons is achieved through the
interaction of the photons with the non-linear media of a
microcavity, filled by a dye solution. While the main contribution
to the interaction in the experiment, reported in
Ref.~\onlinecite{Klaers_Nature}, is thermooptic, it is not a contact
interaction.  It is known that BEC for bosons can exist without
particle-particle interactions~\cite{Einstein} (see
Ref.~\onlinecite{Pitaevskii} for the details), but at least the
interactions with the surrounding media are necessary to achieve
thermodynamical equilibrium. For photon BEC it can be achieved by
interaction with incoherent phonons~\cite{Snoke_Girvin}.   The
influence of interactions on condensate-number fluctuations in a BEC
of microcavity photons was studied in Ref.~\onlinecite{Stoof}.  The
kinetics of photon thermalization and condensation was analyzed in
Refs.~\onlinecite{Kirton1,Kirton2,Kirton3}.   The kinetics of
trapped photon gas in a microcavity, filled by a  dye solution,  was
studied, and, a crossover between driven-dissipative system laser
dynamics and a
thermalized Bose-Einstein condensation of photons was observed~\cite%
{Klaers_2015}.

In previous theoretical studies the equation of motion for a BEC of
photons confined by the axially symmetrical trap in a microcavity
was obtained. It was assumed that the changes of the cavity width
are much smaller than the width of the trap~\cite{Szymanska}. This
assumption results in the coordinate-independent effective photon
mass $m_{ph}$ and photon-photon coupling parameter $g$. In this
Paper, we study the local superfluid and normal density profiles for
trapped two-dimensional gas of photons with the coordinate-dependent
effective mass  and  photon-photon coupling parameter in a an
optical microcavity, filled by a  dye solution. We propose the
approach to study the local BEC and local superfluidity of cavity
photon gas in the framework of local density approximation (LDA) in
the traps of larger size without the assumption, that total changes
of the cavity width are much smaller than the size of the trap. In
this case, we study the effects of coordinate-dependent effective
mass  and photon-photon coupling parameter on the   superfluid and
normal density profiles as well as the profiles of the local
temperature of the phase transition for trapped cavity photons. Such
approach  is useful for the mirrors of smaller radius with the high
trapping frequency, which provide BEC and superfluidity for smaller
critical number of photons at the same temperature.

The paper is organized in the following way. In Sec.~\ref{profile},
we obtain the condensate density profile for trapped microcavity
photon BEC with  locally variable mass  and interactions.
 The expression for the number of particles in a
condensate is analyzed in Sec.~\ref{numsec}. In Sec.~\ref{geomsec},
the dependence of the condensate parameters on the geometry of the
trap is discussed. In Sec.~\ref{sup}, we study the collective
excitation spectrum and superfluidity of 2D weakly-interacting Bose
gas of cavity photons. The  results of our calculations are
discussed in Sec.~\ref{res}.  The proposed experiment for measuring
the distribution of the local density of a photon BEC is described
in Sec.~\ref{expsec}. The conclusions follow in Sec.~\ref{conc}.

\section{The condensate density profile}

\label{profile}

While at finite temperatures there is no true BEC in any infinite
untrapped two-dimensional (2D) system, a true 2D BEC quantum phase
transition can be obtained in the presence of a confining
potential~\cite{Bagnato,Nozieres}. In an infinite translationally
invariant two-dimensional system, without a trap, superfluidity
occurs via a Kosterlitz$-$Thouless superfluid (KTS) phase
transition~\cite{Kosterlitz}.  While KTS phase
transition occurs in systems, characterized by thermal equilibrium,
 it survives in a dissipative highly nonequilibrium
system driven into a steady state~\cite{Dagvadorj}.

 The trap for the cavity photons can be formed by the
concave spherical  mirrors of the microcavity, that provide the
axial symmetry for a trapped gas of photons. Thus the transverse
(along $xy$ plane of the cavity)  confinement of photons can be
achieved by using an optical microcavity with a variable width. Let
us introduce the frame of reference, where $z-$axis is directed
along the axis of cavity mirrors, and $(x,y)$ plane is perpendicular
to this axis. The energy spectrum $E(k)$ for small wave vectors $k$
of photons, confined in $z$ direction in an ideal microcavity with
the coordinate-dependent width $L(\mathbf{r})$, is given
by~\cite{Klaers_Nature}
%%%%%%%%%%%%%%%%%%%%%%%%%%%%%%%%%%%%%%%%%%%%%%%%%%%%%%%%%%%%%%%%%%%%%%%%%%%%%%%%%%%%%%%%%%%%%%%%%%%%%
\begin{eqnarray}
\label{Esp} E(k) = \frac{\hbar c}{\sqrt{\varepsilon}}
\sqrt{k_{z}^{2}+ k_{\bot}^{2}} \approx \frac{\hbar \pi c
\tilde{n}}{\sqrt{\varepsilon}L(\mathbf{r})} + \frac{\hbar^{2}
k_{\bot}^{2}}{2 m_{ph}(\mathbf{r})}
 \ ,
\end{eqnarray}
%%%%%%%%%%%%%%%%%%%%%%%%%%%%%%%%%%%%%%%%%%%%%%%%%%%%%%%%%%%%%%%%%%%%%%%%%%%%%%%%%%%%%%%%%%%%%%%%%%%%%
where $m_{ph}(\mathbf{r}) = \pi \hbar
\tilde{n}\sqrt{\varepsilon}/\left[L(\mathbf{r})c\right]$ is
 the transverse (along $xy$ plane of the cavity)
effective coordinate-dependent photon mass, $k_{z}$ is the wave
vector component in $z$ direction along the axis of the cavity, and
$k_{\bot}$ is the wave vector component in $(x,y)$ plane,
perpendicular to the axis of the cavity,    $c$ is the speed of
light in vacuum, $\varepsilon$ is the dielectric constant of the
microcavity, and $\tilde{n} = 1, 2, \ldots$.  Below we consider the
lowest mode $\tilde{n} = 1$. Assuming the energy difference between
the quantization levels $\tilde{n}$, caused by the boundaries of the
cavity, is much larger than the second term in Eq.~(\ref{Esp}) for
the photon energy, corresponding to a single quantum level, for
small $k_{\bot}$, we treat our system  as a quasi-two-dimensional
system in the $(x,y)$ plane.    Let us mention that Eq.~(\ref{Esp})
is valid only if the radius of the curvature of the mirrors is  much
larger than all other length scales in the system under
consideration, i.e.  the healing  length $\xi$ etc., which it true
for the parameters, used for our calculations, as it can be seen
from the estimations below.

 The Gross-Pitaevskii equation for the wave function of
photon condensate in  a weakly-interacting (through a dye solution)
photon gas can be obtained by following to the standard procedure
for derivation of the Gross-Pitaevskii equation for any
weakly-interacting Bose gas~\cite{Pitaevskii} (see also
Ref.~\onlinecite{Elistratov}). Note that the Gross-Pitavevskii-like
equation for the cavity photons  can be obtained also from the
Maxwell equations in a non-linear medium~\cite{Szymanska}. We
generalize the Gross-Pitaevskii equation for cavity photons with the
coordinate dependent mass and photon-photon interaction.
 The energy functional of the trapped 2D
photon BEC in an optical microcavity, filled by a  dye solution,
can be represented as
\begin{eqnarray}
E[\psi] = \int \left[
    - {\hbar c \over 2 \pi \sqrt{\varepsilon}} \psi^* (\mathbf{r}) \nabla_{\mathbf{r}} \left[
    \left( L(\mathbf{r})\nabla_{\mathbf{r}} \psi(\mathbf{r}) \right) \right] +  \left(\pi \hbar c   {1 \over \sqrt{\varepsilon} L(\mathbf{r})} - \mu   \right)
    | \psi(\mathbf{r})|^2 +  {g(\mathbf{r}) \over 2} |\psi(\mathbf{r})|^4  \right] d\mathbf{r}   \label{energyfunct}
    \ ,
\end{eqnarray}
where $\psi(\mathbf{r})$ is the wave function of the photon
condensate,  $\mu$ is the chemical potential of the cavity photons,
determined by the laser pumping, $g(\mathbf{r})$ is the
photon-photon coupling parameter, corresponding to the photon-photon
interaction through the dye molecule. The form of the functional, presented in Eq.~(\ref{energyfunct}), implies  essential dependence
of the microcavity width $L(\mathbf{r})$ as a function of $r$. This leads to the effect of spatially variable effective mass  $m_{ph}(\mathbf{r}) = \pi \hbar
\tilde{n}\sqrt{\varepsilon}/\left[L(\mathbf{r})c\right]$. Besides, the dependence  $L(\mathbf{r})$ leads to the confining potential
 $V(\mathbf{r}) = \pi \hbar c  /\left(\sqrt{\varepsilon} L(\mathbf{r})\right)$, and also to the spatial dependence of the photon-photon coupling parameter $g(\mathbf{r})$.
The energy functional similar to Eq.~(\ref{energyfunct}) was used
for the particle with the coordinate-dependent mass in
Ref.~\onlinecite{Gevorkyan}.  Let us mention that the energy
functional, presented by Eq.~(\ref{energyfunct}), contains the
coordinate-dependent coupling parameter $g(\mathbf{r})$, which was
not under consideration in Ref.~\onlinecite{Gevorkyan}.

Let us mention that taking into account pumping and losses  of
photons in a microcavity  and assuming that they are not very large,
it is reasonable to expect that the spatial distribution of photons
will not differ much (at least, qualitatively)   from the spatial
distribution of photons, obtained from the minimum of the energy
functional~(\ref{energyfunct}),   for the system of the photons in
the thermodynamical equilibrium.  As it is shown from the numerical
solution of the generalized Gross-Pitaevskii equation with pumping
and losses in  Refs.~\onlinecite{Berloff,BKKL}, small pumping and
losses for condensates of bosons lead to only non-essential
quantitative change of the distribution of condensate profile in the
trap. In the opposite case, the large pumping and losses exceeding
some critical values result in quantum turbulence phenomena in the
system, which manifest themselves in breaking continuous condensate
density distribution and pattern formation~\cite{Berloff,BKKL}.

The variation of the energy functional (\ref{energyfunct}) with
respect to $\psi^{*} (\mathbf{r})$ gives the following equation for
the wave function $\psi (\mathbf{r})$ of the 2D photon condensate in
microcavity traps:
\begin{eqnarray}
\label{genereq}
    - {\hbar c \over 2 \pi \sqrt{\varepsilon}}  \nabla_{\mathbf{r}} \left[
    \left( L(\mathbf{r})\nabla_{\mathbf{r}} \psi (\mathbf{r}) \right) \right] +  \left(\pi \hbar c   {1 \over \sqrt{\varepsilon} L(\mathbf{r})} - \mu   \right)
     \psi (\mathbf{r}) + {g (\mathbf{r})\over 2} |\psi (\mathbf{r})|^2 \psi (\mathbf{r}) = 0
    \ .
\end{eqnarray}

 Following the procedure of the
derivation of the Gross-Pitaevskii equation for a cavity photon BEC,
one obtains photon-photon coupling parameter $g(\mathbf{r})$. The
coordinate-dependent photon-photon coupling parameter
$g(\mathbf{r})$ is given by
%%%%%%%%%%%%%%%%%%%%%%%%%%%%%%%%%%%%%%%%%%%%%%%%%%%%%%%%%%%%%%%%%%%%%%%%%%%%%%%%%%%%%%%%%%%%%%%%%%%%%
\begin{eqnarray}
\label{gdef} g(\mathbf{r}) = \frac{\pi \hbar c
A}{\sqrt{\varepsilon}L(\mathbf{r})}
 \ ,
\end{eqnarray}
%%%%%%%%%%%%%%%%%%%%%%%%%%%%%%%%%%%%%%%%%%%%%%%%%%%%%%%%%%%%%%%%%%%%%%%%%%%%%%%%%%%%%%%%%%%%%%%%%%%%%
where the parameter $A$  determines the strength of photon-photon
coupling  and depends on the properties of the medium, through which
the photons interact (see also Ref.~\onlinecite{Szymanska}).

Due to the mirrors' axial symmetry, the wave function  $\psi
(\mathbf{r})$  of the 2D photon condensate has the axial symmetry.
Since in the stationary nonrotating BEC the angular momentum equals
to zero, one can rewrite Eq.~(\ref{genereq}) in the polar
coordinates as
%-------------------------------------------------------------------------
\begin{eqnarray}
\label{ggpe} - \frac{\hbar c }{ 2 \pi\sqrt{\varepsilon}}
    \left( L(r) {d^2 \over d r^2} + {d L(r) \over d r}
    {d \over d r} + {L(r) \over r} {d \over d r}
    \right) \varphi  (r) +  \left(\pi \hbar c   {1 \over \sqrt{\varepsilon} L(r)} -\mu \right)
    \varphi (r) +  {g (r)\over 2} |\varphi (r)|^2 \varphi (r) = 0     \ ,
\end{eqnarray}
%-------------------------------------------------------------------------
where $\varphi  (r)$  is the radial component of
 the condensate wave function $\psi(\mathbf{r})$.

 We consider a trapped Bose gas with a fixed number of particles in a condensate.
 In general case, the density profile for photon BEC
$n(r) = |\varphi(r)|^{2}$ can be obtained by numerical solution of
Eq.~(\ref{ggpe}) for a given function $L(r)$. In this paper, we
focus on obtaining an analytical expression for the profile of the
microcavity photon BEC.
  For a large number of condensate photons~\cite{Stoof},
one can use for this profile  the Thomas$-$Fermi
approximation~\cite{Pitaevskii} analogously to applicability of the
Thomas$-$Fermi approximation for other physical realizations for
bosons. The conditions of applicability of Thomas$-$Fermi
approximation are discussed  in Appendix~\ref{app:A}. In the
Thomas$-$Fermi approximation~\cite{Pitaevskii}, neglecting the
gradient terms acting on the condensate wave function in
Eq.~(\ref{ggpe}), and, assuming the slowly-varying width of the
cavity, for the chemical potential $\mu$, satisfying to the
following condition:
 %-------------------------------------------------------------------------
\begin{eqnarray}
\label{mucond} \mu > \frac{\pi \hbar c}{\sqrt{\varepsilon}L(r)} \ .
\end{eqnarray}
%-------------------------------------------------------------------------
  one gets
 %-------------------------------------------------------------------------
\begin{eqnarray}
\label{ggpe2} \pi \hbar c   {1 \over \sqrt{\varepsilon}L(r)} -\mu
   +  {g (r) \over 2} n^{2}(r)  = 0     \ ,
\end{eqnarray}
%-------------------------------------------------------------------------
where $n(r) = |\psi(\mathbf{r})|^{2}$  is the condensate density. As
follows from Eq.~(\ref{ggpe2}), $n(r)$ is a slowly varying function,
defined by  the slowly varying function $L(r)$:
 %-------------------------------------------------------------------------
\begin{eqnarray}
\label{ggpe3} n(r) = \frac{2\left(\mu - \pi \hbar
c/\left[\sqrt{\varepsilon} L(r)\right]\right)}{g(r)} = \frac{2\mu -
2 m_{ph}(r) c^{2}/\varepsilon}{g(r)} \ .
\end{eqnarray}
%-------------------------------------------------------------------------

Therefore, the photonic BEC exists only if the laser pumping
provides the chemical potential $\mu$ that satisfies  the condition
given by Eq.~(\ref{mucond}). For an axially symmetrical trap
 the maximal radius  of the BEC spot $r_{0}$ is defined  by  $\mu = \pi \hbar
 c/\sqrt{\varepsilon}L(r_{0})$. The corresponding value of
 $r_{0}$ is valid at $r_{0}<R$, where $R$ is the radius of the
 trap, which is defined by the shape of the mirrors.

\section{The number of particles in a condensate}

\label{numsec}

We assume that the axially symmetrical trap has a harmonic shape. As
it is shown in  Ref.~\onlinecite{Klaers_Nf}, when the distance from
the axis of the microcavity to the mirror is essentially smaller
than the radius of the spherical mirror, the harmonic approximation
for the shape of mirrors forming a microcavity is valid. Let us
derive the expression for the chemical potential $\mu$ for the
harmonic trap. For the harmonic trap $V_{0} + \gamma r^{2}/2 = \pi
\hbar c/\left[\sqrt{\varepsilon}L(r)\right]$ we have $L(r) = \pi
\hbar c/\left[\sqrt{\varepsilon}\left(V_{0} + \gamma
r^{2}/2\right)\right]$, where $V_{0} = \pi \hbar
c/\left[\sqrt{\varepsilon}L(r=0)\right]$,  and $\gamma$ is
 the  constant, determining the curvature of the
harmonic trap.  In the approximation, applied for the
 harmonic trap,  $\gamma$ is  defined as $\gamma = m_{1}\Omega^{2}$,
where $m_{1} = 6.7 \times 10^{-36} \ \mathrm{kg}$ is the cavity
effective photon mass  at $r=0$, given
 in Ref.~\onlinecite{natcom}, and  $\Omega$ is the  frequency of the harmonic trap,
defined as $\Omega =  c \sqrt{2/\left(L(r=0)
R_{m}\right)}$~\cite{Klaers_Nature}.  The radius of the curvature of
the mirrors $R_{m}$ is related to the trapping frequency $\Omega$
as~\cite{Klaers_Nature}
%-------------------------------------------------------------------------
\begin{eqnarray}
\label{trfr} R_{m} = \frac{2c^{2}}{L(r=0)\Omega^{2}} \ .
\end{eqnarray}
%-------------------------------------------------------------------------
 Therefore, for the harmonic trap, one obtains
%-------------------------------------------------------------------------
\begin{eqnarray}
\label{mphharm} m_{ph} (r) = \frac{\varepsilon}{c^{2}} \left(V_{0} +
\frac{\gamma r^{2}}{2}\right) \  ,
\end{eqnarray}
%-------------------------------------------------------------------------
and
%-------------------------------------------------------------------------
\begin{eqnarray}
\label{gharm} g(r) =  A \left(V_{0} + \frac{\gamma r^{2}}{2}\right)
\ .
\end{eqnarray}
%-------------------------------------------------------------------------

For the density profile of the photonic BEC at $r<r_{0}$ from
Eq.~(\ref{ggpe3}) one gets:
 %-------------------------------------------------------------------------
\begin{eqnarray}
\label{ggpe3harm} n(r) = \frac{2}{A} \left(\frac{\mu}{V_{0} + \gamma
r^{2}/2} - 1 \right)   \ .
\end{eqnarray}
%-------------------------------------------------------------------------
The normalization condition for the density  profile of the photonic
BEC $ N_{BEC} = \int n(r) d \mathbf{r}$ and Eq.~(\ref{ggpe3harm})
lead to
%-------------------------------------------------------------------------
\begin{eqnarray}
\label{normaliz1} N_{BEC} = \frac{4\pi\mu}{A\gamma} \ln \left[ 1+
\frac{\gamma r_{0}^{2}}{2V_{0}}\right] - \frac{2\pi r_{0}^{2}}{A} \
.
\end{eqnarray}
%-------------------------------------------------------------------------
We find the equation for $r_{0}$ using
Eq.~(\ref{ggpe3harm}) from the condition $n(r_{0})=0$:
%-------------------------------------------------------------------------
\begin{eqnarray}
\label{mur0} \mu = V_{0} + \frac{\gamma r_{0}^{2}}{2}
 \ ,
\end{eqnarray}
%-------------------------------------------------------------------------
which results in
%-------------------------------------------------------------------------
\begin{eqnarray}
\label{r0} r_{0} = \sqrt{\frac{2\left(\mu - V_{0}\right)}{\gamma}} \
.
\end{eqnarray}
%-------------------------------------------------------------------------

For the harmonic trap, the chemical potential $\mu$, corresponding
to the existence of the BEC, has to satisfy to the condition, which
follows from Eq.~(\ref{mucond}):
%-------------------------------------------------------------------------
\begin{eqnarray}
\label{mur0harm} \mu \geq  V_{0} + \frac{\gamma r^{2}}{2}
 \ ,
\end{eqnarray}
%-------------------------------------------------------------------------
where $r\leq r_{0}$ (see Eq.~(\ref{mur0})).

Substituting Eq.~(\ref{r0}) into Eq.~(\ref{normaliz1}), we get
 the following expression, which connects the number of
photons in BEC $N_{BEC}$ with  the chemical potential $\mu$ and
other parameters of the system:
%-------------------------------------------------------------------------
\begin{eqnarray}
\label{muexp} N_{BEC} = \frac{4\pi\mu}{A\gamma} \ln
\left[\frac{\mu}{V_{0}}\right]  - \frac{4\pi\left(\mu -
V_{0}\right)}{A\gamma} = \frac{4\pi\mu}{A\gamma} \left[\ln
\left[\frac{\mu}{V_{0}}\right]  + \frac{V_{0}}{\mu} - 1 \right] \ .
\end{eqnarray}
%-------------------------------------------------------------------------
 Eq.~(\ref{muexp}) will be used to calculate the spatial
condensate density profile, applying Eq.~(\ref{ggpe3harm}) (see
Sec.~\ref{res}).

\section{Dependence of the condensate parameters on the geometry of the trap}

\label{geomsec}

 The radius of the BEC spot $r_{0}$  is determined by the number of photons in BEC $N_{BEC}$ and
the parameter of strength of  photon-photon coupling $A$. However,
since the photon-photon interaction strength is currently not well
known, we obtain the parameter of strength of photon-photon coupling
$A$, using the experimental results~\cite{Klaers_Nature}. The
experiment~\cite{Klaers_Nature} was maintained at the finite
temperature $T= 300 \ \mathrm{K}$, when the radius of the photon
spot is different from the radius of BEC spot, while at $T =0 \
\mathrm{K}$, assuming almost all photons belong to BEC, the radii of
the photon spot and BEC spot are equal.  For our calculations in the
framework of the Thomas$-$Fermi approximation at $T= 0 \
\mathrm{K}$, we assume that the radius of the photon spot $r_{0}$
corresponds by the order of magnitude to  the one, reported in the
experiment~\cite{arxiv}. The parameter of strength of photon-photon
coupling $A$  was estimated by substituting $r_{0} = 20 \
\mathrm{\mu m}$, and  $L(r=0) =  1.7 \times 10^{-6} \ \mathrm{m}$,
$\gamma = 7.929 \times 10^{-13} \ \mathrm{J/m^{2}}$, $N_{BEC} = 1.7
\times 10^{5}$, $\varepsilon = 2.045$ from Ref.~\onlinecite{natcom}
into Eqs.~(\ref{mur0}) and~(\ref{muexp}).  Then, one obtains $A =
2.87 \times 10^{-5} \ \mathrm{\mu m^{2}}$.

 According to the experiment, at $T = 300 \ \mathrm{K}$
BEC exists at $N_{BEC}
> N_{c}$, where $N_{c} = 8.5 \times 10^{4}$~\cite{arxiv}.  The
experiments have been performed for $N_{BEC}$ in the range from $3
\times 10^{4}$ up to $5.5 \times 10^{5}$~\cite{natcom}. For our
calculations we use $N_{BEC} = 1.7 \times 10^{5}$. While for the
experimental parameters~\cite{natcom}, implying $\Omega =
 2 \pi \times 36.5 \ \mathrm{GHz}$ and  $\gamma = 3.524
\times 10^{-13} \  \mathrm{J/m^{2}}$, the Thomas$-$Fermi
approximation is not applicable, we use for our calculations $\Omega
=  2 \pi \times 54.75 \ \mathrm{GHz}$ and $\gamma = 7.929 \times
10^{-13}  \ \mathrm{J/m^{2}}$,  where the Thomas$-$Fermi
approximation is valid, as it is demonstrated  in
Appendix~\ref{app:A}.

Let us mention that using $\Omega =  2 \pi \times 54.75 \
\mathrm{GHz}$ larger than the value used in
Ref.~\onlinecite{Klaers_Nature} corresponds to slightly smaller
radius $R_{m}$ of the curvature of the mirrors than in
Ref.~\onlinecite{Klaers_Nature}. Thus, for the parameters, used for
our calculations, we have $R_{m} =  0.444  \ \mathrm{m}$.    In
Ref.~\onlinecite{natcom},
 the mirrors of the radius of curvature $R_{m} = 1 \ \mathrm{m}$  have been used.
  The advantage of using the mirrors of smaller radius
  with higher trapping frequencies  is the increase of the  constant  $\gamma$,
   which results in  higher critical temperature of BEC for the same number of
   photons,  because the critical temperature $T_{BEC}^{(0)}$ of BEC for a
non-interacting Bose gas can be qualitatively estimated as (see,
e.g., Ref.~\onlinecite{Bagnato})
%-------------------------------------------------------------------------
\begin{eqnarray}
\label{estidtc} T_{BEC}^{(0)} \sim \frac{\hbar}{\pi
k_{B}}\left[\frac{6 \gamma N_{BEC}}{m_{ph}(r=0)}\right]^{1/2} \ ,
\end{eqnarray}
%-------------------------------------------------------------------------
where $k_{B}$ is Boltzmann constant. Thus at fixed temperature $T$
the critical number of photons, corresponding to the BEC transition,
is inversely proportional to the  constant $\gamma$. Using
Eq.~(\ref{estidtc}), one obtains at $T = 300 \ \mathrm{K}$, for the
experimental constant  $\gamma = 3.524 \times 10^{-13} \
\mathrm{J/m^{2}}$, the estimation for the critical number of photons
for the BEC transitions as $N_{c} \sim 6.694 \times  10^{3}$. At the
same parameters,  for  used in our calculations  the  constant
$\gamma = 7.929 \times 10^{-13}  \ \mathrm{J/m^{2}}$, the critical
number of photons for the BEC transitions can be estimated as $N_{c}
\sim 2.975 \times  10^{3}$.  The latter demonstrates that for the
microcavity  with smaller radius of the mirrors, implying according
to Eq.~(\ref{trfr}) larger $\Omega$, and, therefore, larger constant
$\gamma$, BEC can be achieved for smaller critical number of photons
at the same temperature. Let us mention that using the mirrors of
the radius $R_{m} = 0.444 \ \mathrm{m}$, which corresponds to used
in our calculations the constant $\gamma = 7.929 \times 10^{-13} \
\mathrm{J/m^{2}}$, does not break the validity of Eq.~(\ref{Esp}),
since this radius of the curvature of the mirrors is  much larger
than all other length scales in the system under consideration.

 Let us mention that the effect of taking into account
the spatial dependence of the cavity effective photon mass $m_{ph}
(r)$ and the photon-photon coupling parameter $g(r)$ can be
illustrated  by the following ratios, calculated with the
parameters, introduced above: $m_{ph}(r=r_{0})/m_{ph}(r=0) =
g(r=r_{0})/g(r=0) = 1.004$. At the location of the condensate  the
change of the cavity width can be illustrated by the ratio:
$L(r=r_{0})/L(r=0) = 0.996$. For the relatively small radius of the
mirror $R_{m} = 0.035 \ \mathrm{m}$, implying $\Omega =  2 \pi
\times  277.9   \ \mathrm{GHz}$ and $\gamma = 2.043 \times 10^{-11}
\ \mathrm{J/m^{2}}$, one obtains $m_{ph}(r=r_{0})/m_{ph}(r=0) =
g(r=r_{0})/g(r=0) = 1.1$.  In this case of the mirrors of such small
radius, at  the location of the condensate the change of the cavity
width can be illustrated by the ratio: $L(r=r_{0})/L(r=0) = 0.909$.
Therefore, for the mirrors of the smaller radius, the coordinate
dependence of the effective photon mass and the photon-photon
coupling parameter is stronger. Note that formation of the traps for
the cavity photons, located in the convexities of this small radius
on a plane mirror, seems to be possible. Besides in this paper we
are interested in small radius of mirrors, because it corresponds to
smaller critical number of photons for BEC at fixed temperature and
microscopical traps can be used for quantum technology applications.

\begin{figure}[tbp]
\includegraphics[width=6.0cm]{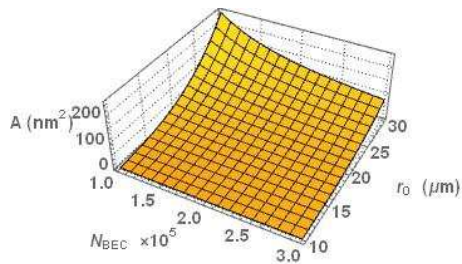}
%\vspace{-10.5 cm}
\caption{The parameter of strength of
photon-photon coupling $A$ as a function of the total number of
photons in the BEC $N_{BEC}$ and the radius of the photon spot
$r_{0}$. } \label{conANr0}
\end{figure}

According to Eq.~(\ref{muexp}), the radius of the BEC spot $r_{0}$ depends  on the parameter of strength of  photon-photon coupling $A$ and the number of photons in BEC
$N_{BEC}$.  We study the  photon-photon interaction strength, assuming the substitution of the experimental values $r_{0}$ and $N_{BEC}$ into Eq.~(\ref{muexp}).
The parameter of strength of photon-photon coupling $A$, required to achieve the certain radius of the BEC spot $r_{0}$ at the defined
total number of photons in the BEC $N_{BEC}$, is shown in Fig.~\ref{conANr0} for various $r_{0}$ and $N_{BEC}$. According to Fig.~\ref{conANr0} at
 the fixed $N_{BEC}$,
 larger $A$ is required
to achieve larger $r_{0}$, and for larger $N_{BEC}$  smaller $A$ is required to achieve the fixed spot radius $r_{0}$.

 The justification of the Thomas$-$Fermi approximation
and slowly varying cavity width approximation for the parameters,
used in our calculations is discussed in Appendix~\ref{app:A}.

\section{Superfluidity of the microcavity photons}

\label{sup}

Below we study the collective excitation spectrum and superfluidity
of 2D weakly-interacting Bose gas of cavity photons.  While at zero
temperature, the entire system is superfluid, at the non-zero
temperatures below the KTS phase transition temperature, in a 2D
superfluid the normal component appears in the cores of the
vortices, with the superfluid, circulating around these
cores~\cite{Onsager,Feynman,Hohenberg,Kosterlitz}. Below we consider
the axially symmetrical trap, where the size of the condensate  is essentially
larger than the average distance between two vortices.
The maximal density of the vortices is estimated as $n_{v}^{(max)} \lesssim r_{v}^{-2}$, where $r_{v}$ is the size of the core of a vortex. The average distance
 $\xi_{av}$  between the vortices cannot be smaller than the size of the core of a vortex $r_{v}$: $\xi_{av} \gtrsim  r_{v}$
(see, e.g., Ref.~\onlinecite{Larkin}). Since the size of the core of
a vortex is of the order of the magnitude of the healing length
$r_{v} \simeq \xi$~\cite{Voronova}, the  size of the condensate  is
larger than the average distance between two vortices, when the
inequality $\xi < r_{0}$ holds, which  does not contradict to the
condition of validity of the Thomas$-$Fermi approximation, presented
by Eq.~(\ref{ineq1}).  Therefore, one can estimate the local
temperature of Kosterlitz-Thouless phase transition, using the
parameters, obtained from the Thomas-Fermi approximation.

 Now we will analyze the spectrum of the collective
excitations in the superfluid of microcavity photons.
 For small momenta  ($P = \hbar k_{\bot}$)
$P \ll \sqrt{2m_{ph}(r) g(r) n(r)}$ and small temperatures, the
energy spectrum of the quasiparticles $\epsilon (P,r)$ is given
by~\cite{Mullin} $\epsilon (P,r) \approx c_{s}(r)P$, where
$c_{s}(r)$ is the sound velocity in the Popov
approximation~\cite{Griffin}:
 %-------------------------------------------------------------------------
\begin{eqnarray}
\label{csr}
 c_{s}(r) =
 \sqrt{\frac{g(r)n(r)}{m_{ph}(r)}} \ .
\end{eqnarray}
%-------------------------------------------------------------------------

For the harmonic trap, substituting Eqs.~(\ref{mphharm})
and~(\ref{gharm}) into Eq.~(\ref{csr}), one obtains
 %-------------------------------------------------------------------------
\begin{eqnarray}
\label{csrharm}
 c_{s}(r) =
c \sqrt{\frac{A n(r)}{ \varepsilon}} = \frac{c}{\sqrt{\varepsilon}}
\sqrt{2 \left(\frac{\mu}{V_{0} + \gamma r^{2}/2} - 1 \right)  } \ .
\end{eqnarray}
%-------------------------------------------------------------------------

The dilute photon gas in an optical microcavity, filled by a  dye
solution,   forms a 2D weakly interacting gas of bosons with the
pair short-range repulsion, caused by the photon-photon interaction
through the dye molecule. Since the spectrum of a weakly interacting
gas of the cavity photons is a linear sound spectrum, satisfying the
Landau criterium of superfluidity~\cite{Abrikosov}, superfluidity of
the cavity photons can be observed in the trap. Therefore, at small
temperatures there are two components in the trapped gas of cavity
photons: the normal component and the superfluid component. We
obtain the number of photons in the superfluid component as a
function of temperature applying the procedure similar to the one
used for the microcavity exciton polaritons in a 2D trap~\cite{BLS}.
We define the total number of particles in the superfluid component
 $N_{s} \equiv N - N_{n}$, where $N_{n}$ is a total number of
particles in the normal component. $N_{n}$ is defined analogously to
the procedure applied for the definition of the density of the
normal component in the infinite system $n_{n}$~\cite{Abrikosov}
using the isotropy of the trapped cavity photonic gas instead of the
translational symmetry for an infinite system. According to the
Landau theory of quasiparticles, at finite temperatures the
non-interacting quasiparticles, contributing to the normal
component, are characterized by the same energy spectrum as the
weakly-interacting particles at the zero
temperature~\cite{Abrikosov}. The Landau theory of quasiparticles is
valid at low temperatures, when the number of particles in the
normal component is much less than the total number of particles:
$N_{n} \ll N$.  The temperatures when our approach is applicable
must be much smaller than the critical phase transition
temperatures. Therefore, the Landau theory of quasiparticles is
valid at these temperatures.  Our estimations using
Eq.~(\ref{estidtc}) show that the possible transition temperatures
even can exceed room temperatures for realistic experimental
parameters.
 Assuming an axially symmetric 2D trap for
microcavity photons, we imagine that a ``gas of quasiparticles''
rotates in the liquid in the plane perpendicular to the axis of the
trap with some small macroscopic angular velocity $\mathbf{\nu}$. In
this case, the distribution function of a gas of quasiparticles can
be obtained from the distribution function of a gas at rest by
substituting for the energy spectrum of the quasiparticles $\epsilon
(P) - \mathbf{M} \mathbf{\nu}$, where $\mathbf{M} = \mathbf{r}
\times \mathbf{P}$ is the angular momentum of the particle. Assuming
$Pr/\hbar \gg 1$, we apply the quasiclassical approximation for the
angular momentum: $M \approx Pr$ and $\epsilon (M,r) = c_{s}P =
r^{-1}c_{s}(r)M$. The total angular momentum in a trap per unit of
area $\mathbf{M}_{\rm tot}(r)$ is given by
%-------------------------------------------------------------------------
\begin{eqnarray}
\label{L_tot} \mathbf{M}_{\rm tot}(r) = \int\frac{d^{2}M}{(2\pi
\hbar r)^{2}} \mathbf{M} n_{B}\left(\epsilon(r,L) - \mathbf{M}
\mathbf{\nu}\right) \ ,
\end{eqnarray}
%-------------------------------------------------------------------------
where we assume that at small temperatures the quasiparticles are
noninteracting, and they  are described by the Bose-Einstein
distribution function $n_{B}(\epsilon) = (\exp[\epsilon/(k_{B}T)] -
1)^{-1}$. For small angular velocities, $n_{B}\left(\epsilon -
\mathbf{M} \mathbf{\nu}\right)$ can be expanded in terms of
$\mathbf{M} \mathbf{\nu}$. Then  in the linear approximation we get
%-------------------------------------------------------------------------
\begin{eqnarray}
\label{L_tot_1} \mathbf{M}_{tot}(r) = - \int\frac{d^{2}M}{(2\pi
\hbar r)^{2}}  \mathbf{M}(\mathbf{M}\mathbf{\nu})\frac{\partial
n_{B}(\varepsilon)}{\partial\epsilon}  \ .
\end{eqnarray}
%-------------------------------------------------------------------------
Assuming that only quasiparticles contribute to the total angular
momentum, we define the density of the normal component $n_{n}(r,T)$
by $M_{tot}(r) = n_{n}(r,T)M_{0}(r)$, where $M_{0}(r) =
m_{ph}(r)r\nu$ is the angular momentum of one quasiparticle. From Eq.~(\ref{L_tot_1}), the
local coordinate-dependent density of the normal component  is obtained as
%-------------------------------------------------------------------------
\begin{eqnarray}
\label{ln_n}
 n_{n}(r,T) = \frac{3 \zeta (3) k_{B}^{3} T^3}{2 \pi \hbar^{2}
c_{s}^{4}(r)m_{ph}(r)} = \frac{3 \zeta (3) k_{B}^{3}L(r)c T^3}{4
\pi^{2} \hbar^{3}\sqrt{\varepsilon} c_{s}^{4}(r)}\ .
\end{eqnarray}
%-------------------------------------------------------------------------
Let us mention that the density of the normal component
$n_{n}(r,T)$ does not depend on the angular velocity of rotation
$\nu$, because $n_{n}(r,T)$ is a linear response of the total
angular momentum in a trap per unit of area  on the external angular velocity. Hence,
$n_{n}(r,T)$ is determined only by the equilibrium properties of the
system.

The temperature dependence of the local density of the superfluid
component $n_{s} (r,T)$ is given by
%-------------------------------------------------------------------------
\begin{eqnarray}
\label{ln_s1} n_{s} (r,T) = n (r) - n_{n} (r,T) \ ,
\end{eqnarray}
%-------------------------------------------------------------------------
where $n(r)$ is the profile of the total photon density, which
almost does not change at low temperatures. Assuming that at low
temperatures the majority of photons belong to BEC, and substituting
Eqs.~(\ref{ggpe3}) and~(\ref{ln_n}) into Eq.~(\ref{ln_s1}), we
obtain the temperature dependence of the local coordinate-dependent
density of the superfluid component:
%-------------------------------------------------------------------------
\begin{eqnarray}
\label{ln_s}
 n_{s}(r,T)  = \frac{2\left(\mu - \pi \hbar c/ \sqrt{\varepsilon}
L(r)\right)}{g(r)} - \frac{3 \zeta (3) k_{B}^{3}c L(r) T^3}{4
\pi^{2} \hbar^{3} \sqrt{\varepsilon}c_{s}^{4}(r)} \ .
\end{eqnarray}
%-------------------------------------------------------------------------

For the total number of photons in the normal component we obtain
 %-------------------------------------------------------------------------
\begin{eqnarray}
\label{N_n}
 N_{n}(T) = 2\pi \int_{0}^{r_{0}} n_{n}(r,T) r dr =  \int_{0}^{r_{0}}
\frac{3 \zeta (3) k_{B}^{3} T^3}{\hbar^{2} c_{s}^{4}(r)m_{ph}(r)} r
dr \ ,
\end{eqnarray}
%-------------------------------------------------------------------------
 where  $\zeta (z)$ is the Riemann zeta function ($\zeta (3) \simeq 1.202$),
$k_{B}$ is Boltzmann constant, and we assume that at low
temperatures almost all photons are in the condensate.

For the total number of photons in the superfluid  component we get
 %-------------------------------------------------------------------------
\begin{eqnarray}
\label{N_s}
 N_{s}(T) = N - N_{n}(T) =  N - \int_{0}^{r_{0}}
\frac{3 \zeta (3) k_{B}^{3} T^3}{\hbar^{2} c_{s}^{4}(r)m_{ph}(r)} r
dr \ .
\end{eqnarray}
 %-------------------------------------------------------------------------

We assume  that the  width of the cavity $L(r)$ very slowly depends
on the coordinate on scales of the order of the mean separation
between vortexes (but the total change of $L(r)$ in the trap is
essential). The superfluid-normal phase transition in the 2D system
is the Kosterlitz-Thouless transition~\cite{Kosterlitz}, and the
local coordinate-dependent temperature of this transition $T_c $ in
a two-dimensional microcavity photon system is determined by the
 equation~\cite{Kosterlitz}:
%%%%%%%%%%%%%%%%%%%%%%%%%%%%%%%%%%%%%%%%%%%%%%%%%%%%%%%%%%%%%%%%%%%%%%%%%%%%%%%%%%%%%%%%%%%%%%%%%%%%%%%%%%%%%%%%%%%%%%%%%%%%%%%%%%%%%%%%%%%%%%%%%%%%%%%%%%
\begin{eqnarray}  \label{T_KT}
T_c (r) = \frac{\pi \hbar ^2 n_s (r,T_c(r))}{2 k_B m_{ph}(r)} \ .
\end{eqnarray}
%%%%%%%%%%%%%%%%%%%%%%%%%%%%%%%%%%%%%%%%%%%%%%%%%%%%%%%%%%%%%%%%%%%%%%%%%%%%%%%%%%%%%%%%%%%%%%%%%%%%%%%%%%%%%%%%%%%%%%%%%%%%%%%%%%%%%%%%%%%%%%%%%%%%%%%%%%
 We can use Eq.~(\ref{T_KT})  only in the framework of the quasilocal
approximation, assuming very slow changes of $L(r)$, such that the
characteristic length of the changes in $L(r)$ is much less than the
average distance between the vortices in the superfluid.

Substituting Eq.~(\ref{ln_s}) for the density $n_{s}(r,T)$ of the
superfluid component into Eq.~(\ref{T_KT}), we obtain an equation
for the local Kosterlitz-Thouless transition temperature $T_{c}(r)$.
The solution of this equation is
%%%%%%%%%%%%%%%%%%%%%%%%%%%%%%%%%%%%%%%%%%%%%%%%%%%%%%%%%%%%%%%%%%%%%%%%%%%%%%%%%%%%%%%%%%%%%%%%%%%%%%%%%%%%%%%%%%%%%%%%%%%%%%%%%%%%%%%%%%%%%%%%%%%%%%%%%%
\begin{eqnarray}
\label{tct} T_c (r)= \left[\left( 1 +
\sqrt{\frac{32}{27}\left(\frac{m_{ph}(r) k_{B}T_{c}^{0}(r)}{\pi
\hbar^{2} n(r)}\right)^{3} + 1} \right)^{1/3} - \left(
\sqrt{\frac{32}{27} \left(\frac{ m_{ph}(r) k_{B}T_{c}^{0}(r)}{\pi
\hbar^{2} n(r)}\right)^{3} + 1} - 1 \right)^{1/3}\right]
\frac{T_{c}^{0}(r)}{ 2^{1/3}} \  ,
\end{eqnarray}
%%%%%%%%%%%%%%%%%%%%%%%%%%%%%%%%%%%%%%%%%%%%%%%%%%%%%%%%%%%%%%%%%%%%%%%%%%%%%%%%%%%%%%%%%%%%%%%%%%%%%%%%%%%%%%%%%%%%%%%%%%%%%%%%%%%%%%%%%%%%%%%%%%%%%%%%%%
where $T_{c}^{0}(r)$ is the local temperature at which the
superfluid density vanishes in the mean-field approximation at the
points with the coordinate vector $r$ (i.e., $n_{s}(r,T_{c}^{0}(r))
= 0$),
%%%%%%%%%%%%%%%%%%%%%%%%%%%%%%%%%%%%%%%%%%%%%%%%%%%%%%%%%%%%%%%%%%%%%%%%%%%%%%%%%%%%%%%%%%%%%%%%%%%%%%%%%%%%%%%%%%%%%%%%%%%%%%%%%%%%%%%%%%%%%%%%%%%%%%%%%%
\begin{equation}
\label{tct0} T_c^0 (r) = \frac{1}{k_{B}} \left( \frac{ 2 \pi
\hbar^{2} n(r) c_s^4(r)m_{ph}(r) }{3 \zeta (3)} \right)^{1/3} \ .
\end{equation}
%%%%%%%%%%%%%%%%%%%%%%%%%%%%%%%%%%%%%%%%%%%%%%%%%%%%%%%%%%%%%%%%%%%%%%%%%%%%%%%%%%%%%%%%%%%%%%%%%%%%%%%%%%%%%%%%%%%%%%%%%%%%%%%%%%%%%%%%%%%%%%%%%%%%%%%%%%
 Eqs.~(\ref{tct}),~(\ref{tct0})  generalize the results
of Ref.~\onlinecite{BKZ} for the coordinate-dependent particle mass
and photon-photon interaction.

\section{Results and discussion}

\label{res}

\begin{figure}[h]
\includegraphics[width=13.0cm]{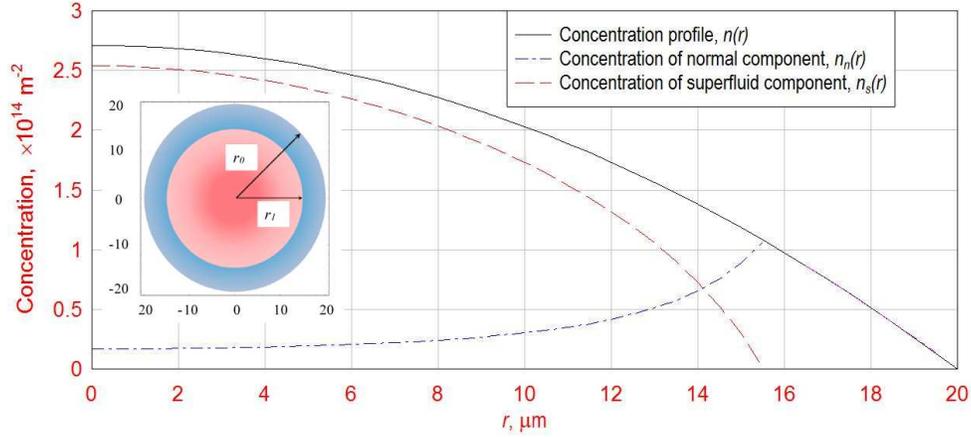}
\caption{The profiles for microcavity photons in a trap for the
total concentration $n(r)$, the concentrations  of the normal
 $n_{n}(r)$ and superfluid
 $n_{s}(r)$ components at the temperature $T = 300 \ \mathrm{K}$.
 The  insert shows the spot of a radius $r_{0}$ formed
by trapped microcavity photons, filled by the superfluid component
in the spot of a radius $r_{1}$, while the normal component is
everywhere inside the spot of a radius $r_{0}$. The gradient of the
colors illustrates the distributions of the concentrations of the
superfluid and normal components.} \label{Fig1}
\end{figure}

Since $T_{c}(r)$ depends on the coordinate $r$, at fixed finite
temperatures $T$ above the minimal possible critical temperature
$T_{c}^{min}=0 \ \mathrm{K}$ at the edge of the BEC  and below the
maximal possible critical temperature $T_{c}^{max}$, e.g.,    $0 \
\mathrm{K} < T < T_{c}^{max}$, there is the superfluid (S) phase in
the region of the system where $T<T_{c}(r)$ (in S phase the
superfluid component coexists with the normal component), and  the
normal (N) phase in the other regions of the system where
$T>T_{c}(r)$ (with only normal component). At the zero temperature,
the entire system is superfluid, and at the temperatures above
$T_{c}^{max}$,
 the entire system is normal. Since  $T_{c}(r)$ is a decreasing
 function of $m_{ph}(r)$, and $m_{ph}(r)$ is a decreasing function
 of the width of the microcavity $L(r)$, $T_{c}(r)$ increases, if
 $L(r)$ increases. If we consider the axially symmetrical trap, where
  $L(r)$ is a decreasing function of
  $r$, then $T_{c}(r)$ decreases, when $r$ increases. Therefore, for
  the axially symmetrical trap we have $T_{c}^{max} = T_{c}(0)\equiv T_{c}(r=0)$. If we consider the temperature $T_{1}$ in the range   $0 \ \mathrm{K} < T_{1} <
  T_{c}(r=0)$, then we have $T_{c}(r_{1}) = T_{1}$, where
  $T_{c}(r_{1})$ is the critical temperature, corresponding to the
  width of the cavity $L (r_{1})$, which can be found from the
  solution of Eq.~(\ref{tct}) with respect to $L(r)$, substituting
  $T_{c} = T_{1}$.   The corresponding  $r_{1}$ is the radius of the spot with the superfluid and normal components  inside and only the normal component outside,
  filling the ring with the
width $r_{0} - r_{1}$  (see the insert in Fig.~\ref{Fig1}).
  While the total density and
  superfluid density are monotonously decreasing functions of $r$ due to the increase of the effective photonic mass with the increase of $r$,
  the  normal density is a non-monotonous function of $r$. At
  $0<r<r_{1}$, $n_{n}(r) = n(r) - n_{s}(r)$ increases with the increase of $r$ due to the
  decrease of $n_{s}(r)$, but at $r_{1}\leq r \leq r_{0}$, $n_{n}(r) =
  n(r)$ is a decreasing function of $r$. The profiles for the
total concentration $n(r)$, the  concentrations of the normal
 $n_{n}(r)$ and  superfluid
 $n_{s}(r)$ components at the temperature $T = 300 \ \mathrm{K}$ are shown  in
 Fig.~\ref{Fig1}. The spatial distributions of the superfluid and normal components in the trap are demonstrated in  the insert in
 Fig.~\ref{Fig1}, where the gradient of the colors
 reflects the local concentrations of the superfluid and normal
 components.

According to Eqs.~(\ref{tct}) and~(\ref{tct0}), the local mean-field phase transition
temperature $T_c^0 (r)$ and local Kosterlitz-Thouless transition
temperature $T_c (r)$ decrease with the increase of the distance
from the center of the trap $r$, and  both $T_c^0 (r)$ and $T_c (r)$
vanish at the edge of the trap. Everywhere inside a trap, the local
mean-field phase transition temperature $T_c^0 (r)$ is greater than
the local Kosterlitz-Thouless transition temperature $T_c (r)$, and
the difference between  $T_c^0 (r)$ and $T_c (r)$ decreases with the
increase of the distance from the center of the trap $r$.

 The superfluidity of cavity photons can be observed
experimentally analogously to the system of microcavity polaritons:
(1) by observing the photon condensate flow induced by initial
gradient of density through the obstacles impurities, where the
superfluid flow does not experience any scattering at the obstacles,
and  (2) by observing the quantized vortices in the system of cavity
photons. The experimental evidence for superfluid motion of exciton
polaritons in semiconductor microcavity was reported in
Ref.~\onlinecite{Amo}. The superfluidity of microcavity exciton
polaritons was studied in terms of the Landau criterion and
manifested itself as the suppression of scattering from defects when
the flow velocity was slower than the speed of sound in the
fluid~\cite{Amo}. We suggest to generalize the methods used to
observe the superfluidity of microcavity polaritons to observe the
superfluidity of the photons in an optical cavity, filled with
molecular medium, that are excited by laser light.

\section{Proposed experiment for measuring the distribution of the local density of a photon BEC}

\label{expsec}

\begin{figure}[h]
\includegraphics[width=8.0cm]{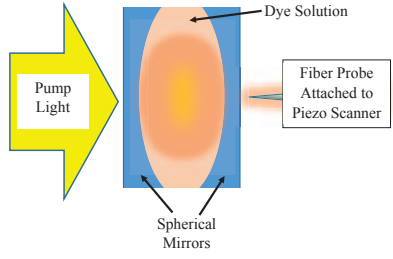}
\vspace{-5cm}
 \caption{Scheme of possible experiment for study of the local density of microcavity photons� condensate. A fiber based
detector is located near the mirrors and can be scanned along the
surface to register the spatial distribution of photons, escaping
the microcavity.} \label{experfig}
\end{figure}

 We propose the following experiment relevant to BEC and
superfluidity of trapped microcavity photons. These experiments are
based on the observation of  local distribution of photons, escaping
the optical microcavity.  The fiber photodetector, formed by a
single fiber probe, attached to a piezo scanner, can be
 moved  above different regions of the mirrors at the distance from
the mirrors about several microns, which is much less than the size
of the BEC. These photodetectors can register the local intensity of
the lines of the angular distribution of light, which is
proportional to the number of the photons, escaped from the nearest
to the detector region of the microcavity with the given angle
$\alpha$  between the momentum of photons escaping the optical
microcavity and the normal to the microcavity. In the absence of
photon flow, the average angle between the momentum of photons
escaping the optical microcavity and the normal to the microcavity
is $\bar{\alpha}=0$, because the angular distribution is
symmetrical. The photons, escaping from
 the BEC inside the circle of the radius $r_{1}$, form the sharp
bright spot with very narrow line, registered by the fiber
photodetector, because the photons from BEC are characterized by
(almost) zero momentum component  $P = 0$ in $(x,y)$ plane, normal
to the axis of the cavity. Hence, these photons from the BEC escape
in the direction normal to the plane of the microcavity. The photons
escape from the non-condensate with various $P$, characterized by
various angles $\alpha$ between the momentum  and the normal to the
microcavity. Hence, photons escape from the non-condensate, forming
line broadening. Therefore, photons escaping from the BEC region
inside the circle of radius $r_{1}$ will form very narrow line of
very high intensity, corresponding to the BEC, and this narrow line
will be surrounded by the broad lower intensity line, corresponding
to the non-condensate. The photons, escaping from the ring of the
inner radius $r_{1}$ and the outer radius $r_{0}$ form only  the
broad lower intensity line, because there is no BEC inside this
ring.  The scheme of this possible experiment is presented in
Fig.~\ref{experfig}. A fiber based detector located near the mirror
can be scanned along the surface to register the spatial
distribution of photons, escaping the microcavity, or the detectors
can be also located inside the microcavity if the corresponding
change of the microcavity quality factor would be negligible. The
quasiparticles, forming the local normal component, contribute to
the local line broadening, which affect the local average deviation
of the tangent of the angle between the path of the escaping photon
and the normal to the microcavity, defined as
%%%%%%%%%%%%%%%%%%%%%%%%%%%%%%%%%%%%%%%%%%%%%%%%%%%%%%%%%%%%%%%%%%%%%%%%%%%%%%%%%%%%%%%%%%%%%%%%%%%%%%%%%%%%%%%%%%%%%%%%%%%%%%%%%%%%%%%%%%%%%%%%%%%%%%%%%%
\begin{equation}
\label{tan} \overline{\Delta \tan \alpha} (r)  = \frac{\sqrt{\overline{P^{2}}(r)}}{p_{z}(r)}  \ .
\end{equation}
%%%%%%%%%%%%%%%%%%%%%%%%%%%%%%%%%%%%%%%%%%%%%%%%%%%%%%%%%%%%%%%%%%%%%%%%%%%%%%%%%%%%%%%%%%%%%%%%%%%%%%%%%%%%%%%%%%%%%%%%%%%%%%%%%%%%%%%%%%%%%%%%%%%%%%%%%%
In Eq.~(\ref{tan}) $p_{z} (r) = \pi\hbar/L(r)$  is the momentum component in $z$ direction along the axis of the
cavity, and the average squared momentum component in $(x,y)$ plane $\overline{P^{2}}$    is given by
%-------------------------------------------------------------------------
\begin{eqnarray}
\label{psq} \overline{P^{2}}(r) =  \frac{1}{n(r)} \int P^{2}
n_{B}(\epsilon)\frac{d^{2}P}{(2\pi
\hbar )^{2}}
\ ,
\end{eqnarray}
%-------------------------------------------------------------------------
where $\epsilon = c_{s}(r)P$ for $0 \leq r <r_{1}$.
In Eq.~(\ref{psq}) it was assumed that the broadening  of the photon angle distribution for a weakly interacting photon gas is
formed only by the contribution of quasiparticles.

For the region of the trap with the superfluid phase ($0 \leq r<r_{1}$) after integration in Eq.~(\ref{psq}) one obtains:
%-------------------------------------------------------------------------
\begin{eqnarray}
\label{psq1} \overline{P^{2}}(r) =
\frac{\Gamma(4) \zeta(4) k_{B}^{4}T^{4}}{2 \pi \hbar^{2} c_{s}^{4}(r) n(r)} \ ,
\end{eqnarray}
%-------------------------------------------------------------------------
where $\Gamma (n)$ is the gamma function  ($\Gamma(4) = 6$) and   $\zeta (z)$ is the Riemann zeta function ($\zeta (4) \simeq 1.0823$).

The broadening of the photon angle distribution for the region  $0 \leq r <r_{1}$ (where the superfluid component exists) is formed by the contribution of quasiparticles
according to Eq.~(\ref{psq1}). Substituting Eq.~(\ref{psq1}) into Eq.~(\ref{tan}), one obtains  the following expression:
%%%%%%%%%%%%%%%%%%%%%%%%%%%%%%%%%%%%%%%%%%%%%%%%%%%%%%%%%%%%%%%%%%%%%%%%%%%%%%%%%%%%%%%%%%%%%%%%%%%%%%%%%%%%%%%%%%%%%%%%%%%%%%%%%%%%%%%%%%%%%%%%%%%%%%%%%%
\begin{equation}
\label{tan2} \overline{\Delta \tan \alpha} (r,T)  = \left(\frac{3\zeta (4)}{\pi n(r)}\right)^{1/2} \frac{k_{B}^{2}T^{2}}{\hbar c_{s}^{2}(r)p_{z}(r)}  \ .
\end{equation}
%%%%%%%%%%%%%%%%%%%%%%%%%%%%%%%%%%%%%%%%%%%%%%%%%%%%%%%%%%%%%%%%%%%%%%%%%%%%%%%%%%%%%%%%%%%%%%%%%%%%%%%%%%%%%%%%%%%%%%%%%%%%%%%%%%%%%%%%%%%%%%%%%%%%%%%%%%
Substituting Eq.~(\ref{tan2})  into Eq.~(\ref{ln_n}), we get
%%%%%%%%%%%%%%%%%%%%%%%%%%%%%%%%%%%%%%%%%%%%%%%%%%%%%%%%%%%%%%%%%%%%%%%%%%%%%%%%%%%%%%%%%%%%%%%%%%%%%%%%%%%%%%%%%%%%%%%%%%%%%%%%%%%%%%%%%%%%%%%%%%%%%%%%%%
\begin{equation}
\label{nntan} n_{n}(r,T) = \frac{\zeta (3)}{4}\left(\frac{3 n(r)}{\zeta (4)\pi \varepsilon}\right)^{1/2} \frac{c k_{B} T}{\hbar c_{s}^{2}(r)} \overline{\Delta \tan \alpha} (r,T)   \ ,
\end{equation}
%%%%%%%%%%%%%%%%%%%%%%%%%%%%%%%%%%%%%%%%%%%%%%%%%%%%%%%%%%%%%%%%%%%%%%%%%%%%%%%%%%%%%%%%%%%%%%%%%%%%%%%%%%%%%%%%%%%%%%%%%%%%%%%%%%%%%%%%%%%%%%%%%%%%%%%%%%
Therefore, one obtains the profile of the density of the normal component $n_{n}(r)$  through an experimental measurement of the profile of
$\overline{\Delta \tan \alpha} (r,T)$.

\begin{figure}[tbp]
\includegraphics[width=8.0cm]{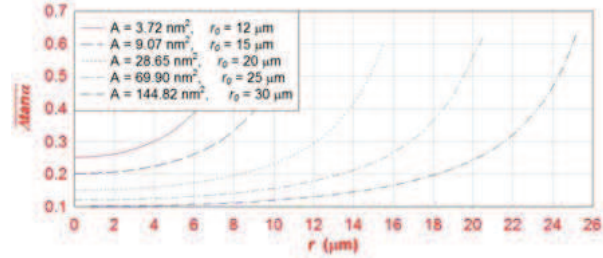}
%\vspace{0 cm}
\caption{The profiles for $\overline{\Delta \tan
\alpha}$ in a trap for $N = 1.7 \times 10^{5}$
  at $T = 300 \ \mathrm{K}$ in the presence of the superfluid
component for $0\leq r < r_{1}$ for different parameters
of strength of photon-photon coupling $A$ and different radii of the photon spot $r_{0}$.} \label{tangraph2D}
\end{figure}

\begin{figure}[tbp]
\includegraphics[width=6.0cm]{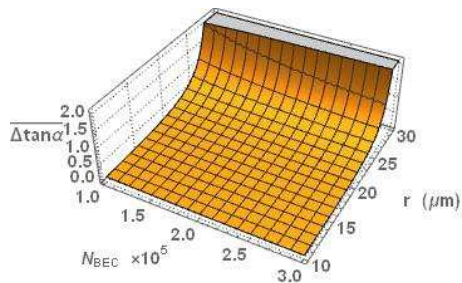}
%\vspace{-12.5 cm}
\caption{The profile for $\overline{\Delta \tan
\alpha}$ in a trap for  $r_{0} = 30 \ \mathrm{\mu m}$ at $T = 300 \
\mathrm{K}$ in the presence of the superfluid component for $0\leq r
< r_{1}$ and different numbers of photons $N$.} \label{tangraph3D}
\end{figure}

The profiles for $\overline{\Delta \tan \alpha}$ in a trap for the total number of photons $N = 10^{5}$ at $T = 300 \ \mathrm{K}$ for the region, where
the superfluid component exists, are presented  in  Fig.~\ref{tangraph2D}
 for different parameters   of strength of photon-photon coupling
$A$ and  radii of the photon spot $r_{0}$.
   According to  Fig.~\ref{tangraph2D},
  $\overline{\Delta \tan \alpha}$  increases up to the edge of the superfluid spot $r_{1}$ and the latter one increases with the  increase of $A$ and $r_{0}$.
   The dependence of the profile of  $\overline{\Delta \tan \alpha}$ on the total number of photons  $N$  and the radius
    of the photon spot $r_{0}$  at $T=300 \ \mathrm{K}$ is presented in Fig.~\ref{tangraph3D}.  In Fig.~\ref{tangraph3D}, $N$ is the total number of photons
    which equals to the number of photons in BEC $N_{BEC}$ at $T = 0 \ \mathrm{K}$, given by Eq.~(\ref{muexp}).
      According to  Fig.~\ref{tangraph3D}, $\overline{\Delta \tan \alpha}$ decreases with the increase of $N$ at the fixed radius of the photon spot
   $r_{0}$.   According to  Figs.~\ref{tangraph2D} and~\ref{tangraph3D},
 $\overline{\Delta \tan \alpha}$ always increases with $r$ at the fixed $r_{0}$, $N$, $A$, and $T$, because the concentration of the normal component always increases
  with $r$ in the region $0\leq r < r_{1}$ according to Fig.~\ref{Fig1}, because only the quasiparticles from the normal component contribute to
  $\overline{\Delta \tan \alpha}$.
  It follows from Figs.~\ref{tangraph2D} and~\ref{tangraph3D}, that by measuring $\overline{\Delta \tan \alpha}$ experimentally, one can obtain
 the parameter of strength of photon-photon coupling $A$, and, therefore, one can
 study photon-photon interaction.

\section{Conclusions}

\label{conc}

In conclusion, we considered the BEC of trapped two-dimensional
 gas of photons with the coordinate-dependent effective mass and   photon-photon coupling parameter in an optical
 microcavity, filled by a  dye solution,
 with the  photons being confined due to the
coordinate-dependent  width of the optical microcavity. The
coordinate dependence of cavity photon effective mass and
photon-photon coupling parameter  describes the photons in a cavity
with the mirrors of smaller radius with the higher trapping
frequency, which provides BEC and superfluidity for smaller critical
number of photons at the same temperature. The photon condensate
density profile was obtained in the  Thomas$-$Fermi approximation.
The condition for the chemical potential, corresponding to the
trapped photonic BEC, was formulated. The local coordinate-dependent
densities of the superfluid and normal components of the trapped
photon system were obtained at a fixed temperature. The profiles of
the superfluid and normal regions were presented at a fixed
temperature.  The profiles for the local mean-field phase transition
temperature $T_c^0 (r)$ and local Kosterlitz-Thouless transition
temperature $T_c (r)$ for trapped microcavity photons  were derived.
The experiments to measure the density profiles for the normal and
superfluid components were suggested.

\acknowledgments

 The authors are thankful to  V. Menon for the
useful discussion. The work was supported by PSC CUNY under Grant
No. 67577-00 45. Yu.~E.~L. was supported by Program of Basic
Research of National Research University HSE.

\appendix

\section{Validity of the Thomas$-$Fermi approximation for the parameters of the calculations}

\label{app:A}

 In this appendix, we discuss validity of the
Thomas$-$Fermi approximation and the condition  of negligibility of
the derivative  $\frac{d L(r)}{d r}$ for the parameters of the
calculations.

 We assume that polariton-polariton interaction is so
week that the mean field approximation is valid.  Local density
approximation (LDA) is applicable when the characteristic condensate
inhomogeneity length, which is the characteristic size of the
condensate $r_0$,  is much larger than all parameters of the problem
with the length dimensionality such as the healing length $\xi$.
When the mean field approximation and the local density
approximation are applicable, then Thomas-Fermi approximation, which
we are using, is valid.

Let us justify the validity of the Thomas$-$Fermi approximation that
is used above. The condition of applicability of the Thomas$-$Fermi
approximation implies neglecting all terms with derivatives in
Eq.~(\ref{ggpe}). Negligibility of these terms, except the term with
$\frac{d L(r)}{d r}$, can be achieved, if the radius of the photon
BEC spot $r_{0}$  is larger than the characteristic length of
decrease of the condensate wave function $\varphi (r)$, which is the
healing length $\xi$. The healing length $\xi$, which corresponds to
the characteristic length of the changes of the condensate wave
function in the Gross-Pitaevskii equation.  The healing length $\xi$
is defined as~\cite{Pitaevskii}
 %-------------------------------------------------------------------------
\begin{eqnarray}
\label{xi} \xi = \frac{\hbar}{\sqrt{2m_{ph}g\bar{n}}}
  \ ,
\end{eqnarray}
%-------------------------------------------------------------------------
where $\bar{n}= N_{BEC}/\left(\pi r_{0}^{2}\right)$ is the average
2D concentration of the photons.

 Besides, negligibility of the term, containing $\frac{d
L(r)}{d r}$, can be achieved at the additional condition, when the
change of the microcavity width $ \Delta L = L(r=0) - L(r=R)$  is
sufficiently smaller than the transverse size of the microcavity
$R$. The Thomas$-$Fermi approximation is applicable if the size of
the condensate $r_{0}$ is much larger than the healing length $\xi$.

Using $N_{BEC} = \pi \bar{n}r_{0}^{2}$, the inequality
 %-------------------------------------------------------------------------
\begin{eqnarray}
\label{ineq1} \xi < r_{0}
\end{eqnarray}
%-------------------------------------------------------------------------
turns into the inequality
%-------------------------------------------------------------------------
\begin{eqnarray}
\label{ineq2} N_{BEC} > \frac{\pi \hbar^{2}}{2m_{ph}g} \ .
\end{eqnarray}
%-------------------------------------------------------------------------

Substituting Eqs.~(\ref{mphharm}) and (\ref{gharm}) into
Eq.~(\ref{ineq2}), and, assuming $r=0$ (which increases the r.h.s.
of Eq.~(\ref{ineq2})), one obtains the following estimate for
$N_{BEC}$ when  the Thomas$-$Fermi approximation is applicable:
%-------------------------------------------------------------------------
\begin{eqnarray}
\label{ineq3} N_{BEC} > \frac{L^{2}(r=0)}{A} \ .
\end{eqnarray}
%-------------------------------------------------------------------------
For the parameters, used for our calculations, from
Eq.~(\ref{ineq3}) one obtains  $N_{BEC} >  10^{5}$. Therefore, for
the used value $N_{BEC} = 1.7 \times 10^{5}$  the Thomas$-$Fermi
approximation can be applied.

Substituting  at $r=0$  Eqs.~(\ref{mphharm}) and~(\ref{gharm})  into
Eq.~(\ref{ineq2}), and assuming $\bar{n} = N_{BEC}/\left(\pi
r_{0}^{2}\right)$, one obtains
 %-------------------------------------------------------------------------
\begin{eqnarray}
\label{xi22} \xi = \frac{r_{0}L(r=0)}{\sqrt{\pi A N_{BEC}}}
  \ ,
\end{eqnarray}
%-------------------------------------------------------------------------
which for the parameters, used in our calculations, results in $\xi
= 8.691 \times 10^{-6} \ \mathrm{m}$. Therefore, for the system
under consideration the inequality $\xi < r_{0}$ holds.

 The another method to check the validity of the
Thomas$-$Fermi approximation is to substitute the condensate wave
function in the form  $\varphi (r) = \varphi_{0}(r) + \delta (r)$
into Eq.~(\ref{ggpe}), where $\varphi_{0}(r)$ is the condensate wave
function in the Thomas$-$Fermi approximation, satisfying
Eq.~(\ref{ggpe2}), and $\delta (r)$ is the small perturbation to the
condensate wave function, caused to the deviation from the
Thomas$-$Fermi approximation.

 Assuming  that the derivatives of
$\delta (r)$ vanish, one obtains
%-------------------------------------------------------------------------
\begin{eqnarray}
\label{fder} F(r)  = - \frac{\hbar c }{ 2 \pi\sqrt{\varepsilon}}
    \left( L(r) {d^2 \over d r^2} + {d L(r) \over d r}
    {d \over d r} + {L(r) \over r} {d \over d r}
    \right) \varphi_{0}  (r) =  - \frac{\hbar c D(r) }{ 2 \pi\sqrt{\varepsilon}}
        \ ,
\end{eqnarray}
%-------------------------------------------------------------------------
where  $\varphi_{0} (r) = \sqrt{n(r)}$, $n(r)$ is given by
Eq.~(\ref{ggpe3harm}), and
%-------------------------------------------------------------------------
\begin{eqnarray}
\label{D123} D(r) = D_{1}(r) + D_{2}(r) + D_{3}(r)  \ ,
\end{eqnarray}
%-------------------------------------------------------------------------
where
%-------------------------------------------------------------------------
\begin{eqnarray}
\label{D1} D_{1}(r) = \frac{L(r)}{2\varphi_{0}(r)}
\left[\frac{d^{2}n(r)}{d r^{2}}  - \frac{1}{2n (r)} \left(\frac{d
n(r)}{d r}\right)^{2}\right] \ ,
\end{eqnarray}
%-------------------------------------------------------------------------
where
%-------------------------------------------------------------------------
\begin{eqnarray}
\label{dndr} \frac{d n(r)}{d r} = - \frac{2\mu \gamma
r}{A\left(V_{0} + \frac{\gamma r^{2}}{2}\right)^{2}} \ ,
\end{eqnarray}
%-------------------------------------------------------------------------
%-------------------------------------------------------------------------
\begin{eqnarray}
\label{d2ndr2} \frac{d^{2} n(r)}{d r^{2}} =  \frac{2\mu \gamma
\left(\frac{3}{2}\gamma r^{2} - V_{0}\right)}{A\left(V_{0} +
\frac{\gamma r^{2}}{2}\right)^{3}} \ ,
\end{eqnarray}
%-------------------------------------------------------------------------
%-------------------------------------------------------------------------
\begin{eqnarray}
\label{D2} D_{2}(r) = \frac{\mu L(r) \gamma^{2} r^{2}}{A
\varphi_{0}(r)
 \left(V_{0} + \frac{\gamma r^{2}}{2}\right)^{3}} \ ,
\end{eqnarray}
%-------------------------------------------------------------------------
%-------------------------------------------------------------------------
\begin{eqnarray}
\label{D3} D_{3}(r) = -\frac{\mu L(r) \gamma}{2A \varphi_{0} (r)
\left(V_{0} + \frac{\gamma r^{2}}{2}\right)^{2}} \ .
\end{eqnarray}
%-------------------------------------------------------------------------

Substituting the expansion $\varphi (r) = \varphi_{0}(r) + \delta
(r)$ and   Eq.~(\ref{fder}) into  Eq.~(\ref{ggpe}), applying
Eq.~Eq.~(\ref{ggpe2}) for $\varphi_{0}(r)$, and keeping only linear
terms with respect to  $\delta (r)$, one obtains  $\delta (r)$ in
the following form
 %-------------------------------------------------------------------------
\begin{eqnarray}
\label{delta}\delta (r) = - \frac{F(r)}{\frac{\pi \hbar c}
{\sqrt{\varepsilon}L(r)} -\mu
  +  {3  \over 2} g(r) n(r)}     \ .
\end{eqnarray}
%-------------------------------------------------------------------------
Substituting the parameters used for our calculations into
Eq.~(\ref{delta}), one obtains $\left|\delta (r=0)/\varphi_{0} (r=0)
\right| = 0.071 \ll 1$. Since it follows from Eq.~(\ref{delta}) that
$\left|\delta (r=0)/\varphi_{0} (r=0) \right| = 0.071 \ll 1$, we
conclude that the Thomas-Fermi approximation is valid for the
parameters, used for our calculations.  At the distance from the
center of the trap $r= 7 \ \mathrm{\mu m}$, this ratio becomes
$\left|\delta (r)/\varphi_{0} (r) \right| = 0.1$. Let us mention
that we present the ratio $\delta/\varphi_{0}$ in the center of the
trap, since closer to the edges of the spot it becomes close to one
due to the well known fact that closer to the edges of the spot the
Thomas$-$Fermi approximation is not valid, and the Gross-Pitaevskii
equation has to be solved~\cite{Mullin}.  While in the framework of
the Thomas$-$Fermi approximation the condensate density vanishes at
the edges of the spot, the solution of the Gross-Pitaevskii equation
demonstrates the asymptotic decrease of the condensate
profile~\cite{Mullin} due to the essential role of the spatial
derivatives of the condensate wave function.

Our assumption about the slowly-varying width of the cavity
corresponds to the inequality  $\Delta L \ll R$, where $\Delta L$ is
the change of the width of the cavity, defined as
%-------------------------------------------------------------------------
\begin{eqnarray}
\label{dell} \Delta L =  L(r=0) - L(r=R) = \frac{\pi \hbar
c}{\sqrt{\varepsilon}V_{0} } - \frac{\pi \hbar
c}{\sqrt{\varepsilon}\left(V_{0} + \gamma R^{2}/2\right)} \ .
\end{eqnarray}
%-------------------------------------------------------------------------
In Eq.~(\ref{dell}) $R$ is the radius  of the microcavity, which is
  $R = 0.5 \ \mathrm{mm}$~\cite{Nyman}.   For the parameters, used for our calculations, Eq.~(\ref{dell})
results in $\Delta L = 1.204 \times 10^{-6} \ \mathrm{m}$.

\end{document}